\documentclass{ptephy}

\usepackage{boites,amsfonts,latexsym,eucal,color,graphicx,epsfig,amssymb,amsmath,booktabs,multirow}

\newcommand{\be}{\begin{eqnarray}}
\newcommand{\ee}{\end{eqnarray}}
\newcommand{\beq}{\begin{eqnarray}}
\newcommand{\eeq}{\end{eqnarray}}
\newcommand{\pd}{\partial}
\newcommand{\nb}{\nabla}

\newcommand{\et}{{\it et al.}}

\begin{document}

\title{Weak cosmic censorship in gravitational collapse with astrophysical parameter values}

\author{\name{\fname{Umpei} \surname{Miyamoto}}{1,\ast}, \name{\fname{Sanjay} \surname{Jhingan}}{2}, and \name{\fname{Tomohiro} \surname{Harada}}{1}\thanks{These authors contributed equally to this work}}

\address{\affil{1}{Department of Physics, Rikkyo University, Tokyo 171-8501, Japan}
\affil{2}{Centre for Theoretical Physics, Jamia Millia Islamia, New Delhi 110025, India}
\email{umpei@rikkyo.ac.jp}}

\begin{abstract}
The possible violation of the weak cosmic censorship hypothesis in astrophysical phenomena can provide us with the information about trans-Planckian physics through observations. We present negative evidence, however, that one should not expect such a possibility at least when the deviation from spherical symmetry is negligible and the parameter values of collapse are astrophysically reasonable. Taking the Lema\^itre-Tolman-Bondi solution as the model most likely to counter the weak hypothesis, we show that the mass ($ \gtrsim 1.5 M_\odot $) and density ($ \gtrsim 1.5 \times 10^{15} \; {\rm g/cm^3}$) of the collapsing object produce a gravitational field strong enough to capture any null rays soon after emanating from the singularity.
\end{abstract}


\maketitle

\section{Introduction}
\label{sec:into}

There are two versions of Cosmic Censorship Hypothesis (CCH)~\cite{Penrose1969}. One is called the weak CCH, stating that spacetime singularities forming in the gravitational collapse of physically reasonable matter that evolves from smooth initial data are hidden behind an event horizon. Another is called the strong CCH, stating that any physical spacetime is globally hyperbolic. Roughly speaking, the former allows a singularity to be locally naked but not globally. The latter does not allow a singularity to be naked even locally.

Since the validity of the CCH or the absence of naked singularities is often assumed to prove theorems in the black-hole physics, the importance of the CCH in general relativity is not to be argued~\cite{Hawking:1973uf}. On the other hand, the possible violation of CCH is also interesting since such violation can provide us with the chance to observationally obtain the information of physics beyond the Planck scale~\cite{Harada:2004mv}.

It is very hard to prove or disprove the CCH in general even in the framework of classical general relativity. It would be still possible and meaningful, however, to study conditions under which the CCH holds or not in our universe, namely, in astrophysically realistic situations. Studies in this context have become an active research area recently. For instance, many researchers are interested in the relevance and consequence of an overspinning Kerr spacetime, which contains a naked singularity. Giacomazzo \et\ \cite{Giacomazzo:2011cv} have recently studied the fate of the collapse of a rapidly differentially rotating neutron star with the angular momentum greater than the Kerr bound and found that generic conditions for such a progenitor do not lead to a naked singularity. In this paper, we are concerned with the CCH in astrophysics, in particular, the visibility of spacetime singularities forming in gravitational collapse.

A natural way to examine the CCH in astrophysically realistic gravitational collapse is to carry out general relativistic numerical simulation, as much microphysics as possible taken into account, and check the causal structure of the obtained spacetime. If one goes to any expense and manages to deal with singularities in simulation, this approach may be the best way at this point. In this paper, however, we take another approach, which is limited in symmetry and matter field but quite simple and straightforward. 

In this paper, we shelve the strong CCH. The validity or invalidity of the strong CCH depends on the local properties just around the singularity, therefore, requiring an accurate description of microphysics (even including quantum gravity) beyond, say, the nuclear density, which is unavailable at this point. Thus, we focus our attention on the weak CCH. That is, we examine the visibility of naked singularity to distant observers.\footnote{We stress that we are not in a position to claim that a singularity being locally naked appears in astrophysically realistic gravitational collapse. Namely, the appearance of such a naked singularity in our analysis is just an assumption to carry on the verification of strong CCH.} In particular, we study the dependence of the causal structure (namely, the behavior of null rays emanating from the singularity) on the mass, density, and initial radius of the collapsing star. The conclusion of this paper is simple. The astrophysical values of the mass, density, and radius for gravitational collapse produce a gravitational field strong enough to trap the null rays soon after emanating from the singularity. Namely, even if {\it by any chance} a naked singularity appears in the gravitational collapse of stellar-size objects, such a naked singularity cannot be visible from the far-away observers, suggesting the existence of a ``weak astrophysical censorship.''

In this paper the above conclusion is drawn from an analysis based on the so-called Lema\^itre-Tolman-Bondi (LTB) solution~\cite{Lemaitre:1933}. These solutions describe the spherically symmetric collapse of a pressureless perfect fluid (\ie, a dust fluid). Here, the readers might wonder if this model could be astrophysically realistic. Again, we need to stress our position here: we consider the LTB solution {\it not} as a model of astrophysically realistic collapse {\it but} as a model which has raised some of the most serious counterexamples to violate the strong CCH.

Here we give some comments on the assumptions of dust fluid and spherical symmetry. Once the dynamical instability sets in and gravitational collapse goes on to some extent, the pressure though being increased no longer holds the collapse but even acts as the source of gravity~\cite{Shapiro:1983du}. In such a situation, as argued in the pioneering paper of Oppenheimer and Snyder~\cite{Oppenheimer:1939ue}, the line element describing the collapse of a pressureless fluid is expected to give a reasonably good approximation to the line element for the collapse of a fluid with nonvanishing pressure. As for the spherical symmetry assumption, it cannot be justified to highly nonspherical collapse. The assumptions of both dust fluid and spherical symmetry may appear to be valid in the late stage of the unbounded core collapse of a very massive star or the delayed collapse of a newly born neutron star if the deviation from spherical symmetry is sufficiently small.

We begin with a quick review of the LTB solution in Sec.~\ref{sec:metric}. In Sec.~\ref{sec:IC} we set arbitrary functions in the LTB solution to suit for our aim. In Sec.~\ref{sec:geodesic} the null geodesic equation is prepared. Then, the global visibility of the singularity is examined in Sec.~\ref{sec:censor}. Section \ref{sec:conc} is devoted to conclusion. We work in the geometrized unit ($ c=G=1 $), otherwise denoted.

\section{Lema\^itre-Tolman-Bondi (LTB) solution}
\label{sec:LTB}

\subsection{Metric, singularity, and apparent horizon}
\label{sec:metric}

The energy-momentum tensor of a pressureless perfect fluid is written as $T^{\mu\nu} = \rho u^\mu u^\nu$, with the energy density $\rho$ and the normalized velocity field $u^\mu$ ($u^\mu u_\mu = -1$). Solving the Einstein equations with the spherically symmetric ansatz, the Lema\^{i}tre-Tolman-Bondi (LTB) solution is obtained in comoving synchronous coordinates ($u^{\mu} = \delta^{\mu}_t$) as
\begin{align}
        ds^2
        &=
        -dt^2 + \frac{R^{\prime 2}}{1+f(r)} dr^2 + R^2(t,r) d\Omega^2,
\label{metric}
\\
        \dot{R}^2
        &=
        f(r) + \frac{F(r)}{R},
\label{Rdot}
\\
        \rho(t,r)
        &=
        \frac{F^\prime}{ 8\pi R^2 R^\prime },
\label{density}
\end{align}
where $X^\prime=\pd_r X$, $\dot{X}=\pd_t X$, and $d\Omega^2 $ is the
line element of a unit two-sphere. $F(r)$ and $f(r)$ are arbitrary functions, but fixed once the initial distributions of mass and velocity of the
fluid are specified.
The physical interpretation of these functions can be seen from equation~\eqref{Rdot}: the left-hand side is a kinetic energy; the second term on the right-hand side plays a role of potential energy; $f(r)$ is a total energy. Integrating equation~\eqref{Rdot}, we obtain $R=R(t,r)$ in an implicit form
\be
        t - t_s(r)
        =
        - \frac{ R^{3/2} }{ \sqrt{F} }
        G \left( - \frac{ fR }{ F } \right),
\label{integration}
\ee
where $t_s(r)$ is an arbitrary
function and
\be
        G(y)
        =
        \begin{cases}
        {\displaystyle \frac{{\rm Arcsin} \sqrt{y}}{y^{3/2}} - \frac{\sqrt{1-y}}{y} }, & 0<y \leq 1 \\
        {\displaystyle \frac{2}{3}}, & y=0 \\
        {\displaystyle - \frac{{\rm Arcsinh} \sqrt{-y}}{ (-y)^{3/2} } - \frac{\sqrt{1-y}}{y} }, & -\infty \leq y < 0 \\
        \end{cases}
        .
\ee
With the coordinate degrees of freedom available, one can choose initial
time at $t=0$ and fix $ R(0,r)=r $ without any loss of generality. Then,
the expression for the time when a fluid element
with $r=const.$ plunges into the shell-focusing singularity, which is defined by $R=0$, is given by
\be
        t_s(r)
        =
        \frac{r^{3/2}}{\sqrt{F}}G \left( - \frac{ fr }{ F } \right).
\label{sing}
\ee
For spherically symmetric systems the condition for 
an apparent horizon, which corresponds to a two-sphere whose outward normals are null, is given by
\be
	\nb_\mu R \; \nb^\mu R = 0 \quad \Longrightarrow \quad R = F.
\label{appcond}
\ee
We would like to note here that an apparent horizon depends on the
slicing of spacetime. However, we are interested in the issue of global
visibility, and at the surface boundary of collapsing dust ball
the apparent horizon in the present choice of time slicing coincides with the event horizon. Thus the apparent horizon, which is much easier to locate, should serve our purpose here.

In LTB models, from equations (\ref{integration}) and (\ref{appcond}), we have the following relation between the singularity and the apparent horizon (AH) curves
\be
        t_{\rm AH}(r)
        =
        t_s(r) - FG(-f).
\label{AH}
\ee
This equation is crucial in understanding the causal structure of
singularity in LTB models. Since $G$ is a positive function, the positivity of mass $F$ implies $t_{\rm AH}(r) <
t_s(r) $ for the non-central shells $(r>0)$. It is only at the centre
that the regularity condition on the initial data demands the mass
function $F$ should be zero and hence the singularity and the apparent
horizon appear simultaneously, \ie, $t_{\rm AH}(0) = t_{s}(0)$. Thus it
is only the central singularity that can be a naked shell-focusing singularity. In what follows we shall be interested in the central shell-focusing singularity.

\subsection{Initial conditions}
\label{sec:IC}

In this paper, we assume the following simple form of the density profile at the initial time $t=0$,
\be
        \rho(0,r)
        =
        \begin{cases}
        {\displaystyle \rho_c \left[ 1-\left( \frac{r}{L} \right)^n \right]}, & 0 \leq r \leq r_b \\
        0, & r > r_b \\
        \end{cases},
\label{rho}
\ee
where $\rho_c$, $L$, and $r_b$ ($\leq L$) are positive constants, and $n=1$, 2, or 3. Here, $\rho_c$ is the central density; $r=r_b$ gives the boundary of the star; $L$ is a parameter controlling the gradient of the density as well as $n$.\footnote{We will show later that the trapping of null rays emanating from the singularity by the apparent horizon happens deep inside the star. This suggests that the detail of density profile around the surface does not affect our conclusion, and therefore that the simple choice of density profile (\ref{rho}) with a discontinuity at the surface is sufficiently good for the present purpose.} Since $F(r)$ is twice the Misner-Sharp quasi-local mass, the above ansatz of density distribution fixes $F(r)$ as
\be
        \frac{1}{2}F(r)
        =
        4 \pi \int_0^r \rho(0,r) r^2 dr.
\label{MS}
\ee
The LTB spacetime is connected to an outer Schwarzschild spacetime whose mass $M$ is given by
\be
        M = \frac{1}{2}F(r_b).
\label{mass}
\ee
For our purpose, it is convenient to choose $(M,\rho_c,r_b,n)$ as a set
of parameters specifying models. From equations~\eqref{rho}, \eqref{MS}, and \eqref{mass}, the remaining parameter $L$ can be given in terms of these parameters
\be
        L
        =
        r_b
        \left(
                \frac{ 12\pi }{ (n+3)(4\pi-3M/\rho_c r_b^3) }
        \right)^{1/n}.
\label{Lrb}
\ee
Although the set of parameters $(M,\rho_c,r_b,n)$ are almost
independent, there are two constraints among them for $L$ to be real and
satisfy $\rho(0,r)\ge 0$. From equation~\eqref{Lrb} such constraints are found to be
\be
        \left(
                \frac{3M}{4\pi \rho_c}
        \right)^{1/3}
        <
        r_b
        \leq
        \left(
                \frac{n+3}{n} \cdot \frac{3 M }{4 \pi \rho_c}
        \right)^{1/3}.
\label{constraint}
\ee
In the limit where $r_b$ takes the value of the left-hand side, $L$
diverges, corresponding to a homogeneous density profile. The right equality holds when $L=r_b$, corresponding to the density vanishing at $r\to r_{b} - 0$.

In addition to $F(r)$, the LTB solution has another arbitrary function $f(r)$, which is related to the initial velocity distribution of fluid elements $\dot{R}(0,r)$. We assume that all fluid elements are at rest $\dot{R}(0,r)=0$ at the initial time, which is called the momentarily static case. From equation~\eqref{Rdot}, such a case is realized by taking
\be
        f = - \frac{F}{r}.
\label{f}
\ee
This condition is very different from the usual ``marginally bound'' models, where collapsing shells are at rest at spatial infinity, corresponding to vanishing $f$. The momentarily static condition is astrophysically motivated because the collapse to a black-hole formation will begin when a dynamical instability sets in. Such an instability is due to: the electron-capture at a static massive stellar core; the photo-dissociation reactions at a core of very massive stars in the prompt core-collapse scenario; the cooling and mass fall-back at a newly born neutron star in the delayed collapse scenario~\cite{Brown:1993jz}.

\subsection{Null geodesic equation}
\label{sec:geodesic}

We are interested in the null geodesics emanating from the central
singularity, which is characterized by $R=r=0$. In particular, one has
to identify the first radial null geodesic emanating from the
singularity, which is a part of the Cauchy horizon, out of all possible
null geodesics. The central singularity is a singular point on the null
geodesic equation from where an infinite family of  null rays come
out. One needs to parameterize these trajectories appropriately to assign each curve a distinct tangent at the singularity. Such a parametrization is possible in ($R,u$)-coordinates rather than ($t,r$)-coordinates, where
$u:=r^\alpha$ with a constant $\alpha$ ($\geq 1$) to be fixed for each
background spacetime so that the null geodesics have finite tangents at the singularity $(R,u)=(0,0)$~\cite{Joshi:1993zg}.

The equation of trajectory $R=R(u)$ for outgoing radial null geodesics is
\be
        \frac{dR}{du}
        =
        \left(
                1-\frac{ \sqrt{f+\Lambda/X} }{ \sqrt{1+f} }
        \right)
        \frac{H(X,u)}{\alpha},
\label{dRdu}
\ee
where a complete list of variables and functions is
\begin{align}
\begin{split}
        u &= r^\alpha,
\;\;\;
        X = \frac{R}{u},
\\
        H(X,u)
        &=
        (\eta-\beta)X
        +
        \left[
                \Theta - \left( \eta-\frac{3}{2} \beta \right) X^{3/2} G(-PX)
        \right]
        \sqrt{ P+\frac{1}{X} },
\\
        \Theta(r)
        &=
        \frac{ 1+\beta-\eta }{ (1+p)^{1/2} r^{3(\alpha-1)/2} }
        +
        \frac{ (\eta-3\beta/2) G(-p) }{ r^{3(\alpha-1)/2} },
\\
        \eta(r)
        &=
        \frac{rF^\prime}{F},
\;\;\;
        \beta(r)
        =
        \frac{rf^\prime}{f},
\;\;\;
        p(r)
        =
        \frac{rf}{F},
\;\;\;
        P(r)
        =
        p r^{\alpha-1},
\;\;\;
        \Lambda(r)
        =
        \frac{F}{r^\alpha}.
\end{split}
\label{defs}
\end{align}
It is noted that $\beta(r)$ is defined to be zero if $f$ is identically zero.

The singularity can be naked if there exists an outgoing null geodesic emanating from it with a definite positive tangent $X_0 := \lim_{u,R \to 0} R/u = \lim_{u,R \to 0} dR/du$. An algebraic equation for such a positive tangent is obtained by taking the limit $(u,R) \to (0,0)$ in equation \eqref{dRdu},
\be
        X_0
        =
        \left(
                1-
                \frac{ \sqrt{ f_0 + \Lambda_0 /X_0 } }
                         { \sqrt{ 1+f_0 } }
        \right)
        \frac{H(X_0,0)}{\alpha},
\label{rootEq}
\ee
where the subscript 0 denotes the limiting value at $r \to 0$. Thus, if there exists a positive root $X_0$ of equation~\eqref{rootEq} for given initial conditions $F$ and $f$, the null ray emanating from the singularity with the tangent $X_0$ is proved to be (a part of) the Cauchy horizon, and therefore, the spacetime is, at least, locally.

\begin{table}[b!]
\begin{center}
\caption{Input parameters ($\rho_c,M,r_b,n$) and parameters ($L, R_{\rm trap},R_{\rm trap}/R_b$) obtained from the present analysis. $\alpha=5/3$ for $n=1$ and $\alpha=7/3$ for $n=2$.}
\label{tbl:para}
\vspace{1mm}
\setlength{\tabcolsep}{6pt}
\begin{tabular}{ccccrrrrrrrrrrrrr}
\hline\hline
Model & $\rho_c \; [10^{15} {\rm g/cm^3}]$ & $M \; [M_\odot]$  & $r_b \; [{\rm km}]$ & $n$ & $L \; [{\rm km}]$  & $R_{\rm trap}\; [{\rm km}]$ & $R_{\rm trap}/R_b$  \\
\hline
\multirow{2}{*}{A} &  \multirow{2}{*}{5.00} & \multirow{2}{*}{1.50} & \multirow{2}{*}{7.00} & 1 & 8.98  & 0.298 & 0.0648 \\
 &   &   &  & 2 & 7.09  & 0.00170 & 0.000339 \\
\hline
\multirow{2}{*}{B} & \multirow{2}{*}{2.00} & \multirow{2}{*}{2.00} & \multirow{2}{*}{10.5} & 1 & 13.3  & 0.533 & 0.0778 \\
 &   &   &   & 2 & 10.6 & 0.00379  & 0.000502 \\
\hline
\multirow{2}{*}{C} & \multirow{2}{*}{1.50} & \multirow{2}{*}{2.70} & \multirow{2}{*}{12.8} & 1 & 16.2 & 0.144 & 0.0656 \\
 &   &   &  & 2 & 12.9 & 0.00320  & 0.000346 \\
\hline\hline
\end{tabular}
\end{center}
\end{table}

\section{Weak cosmic censorship}
\label{sec:censor}

As mentioned above, the nakedness of the central singularity is determined by examining the null rays locally around the singularity for given initial conditions $F$ and $f$. In our models $F$ and $f$ are characterized by the parameters ($M,\rho_c,r_b,n$). It can be shown that for $n=1$ and $n=2$ the singularity is naked, while for $n=3$ the
singularity can be either naked or censored, depending on $M$, $\rho_{c}$ and $r_{b}$. Since we are interested in the weak CCH, we discuss the cases $n=1$ and $n=2$ in this section. We mention
here that the $n=2$ profile is most physically plausible in the sense that it models the mass distribution of stars in hydrostatic equilibrium. This is because in the presence of pressure $p=p(\rho)$, the pressure gradient can be balanced at the centre with the gravitational force, which is proportional to $r$, only for $n=2$. For $n=1$, the pressure gradient force dominates the gravitational force, while the situation is reversed for $n=3$. The case of $n=3$ is relegated to Appendix~\ref{sec:strong}, which involves the strong CCH as well as the weak CCH.

\subsection{Generic results}
\label{sec:weak}

Since we have a large three-dimensional parameter space of ($M,\rho_c,r_b$) only constrained by \eqref{constraint} even if $n$ is fixed, it is practical to take several sets of numerical values of ($M,\rho_c,r_b$), rather than to survey the whole range of parameters. We adopt three sets of parameters, which we call models A, B, and C (see table~\ref{tbl:para}). Models A, B and C are motivated by marginally stable configurations of neutron stars for an extremely soft, medium and extremely hard equations of state, respectively (see, \eg, \cite{Shapiro:1983du}).

For a given set of ($M,\rho_c,r_b,n$), parameter $L$ is determined
through equation~\eqref{Lrb}. The power $\alpha$ in
equation~\eqref{defs} is fixed so that $\Theta(r)$ has a finite value in
the limit $r \to 0$. The results are $\alpha=5/3$ for $n=1$ and
$\alpha=7/3$ for $n=2$. Solving equation~\eqref{rootEq} with $\alpha$
obtained, the positive root $X_0$ is determined. Then, one is ready to integrate~\eqref{dRdu} to obtain the trajectory of Cauchy horizon $R=R(u)$ with the initial conditions of $R|_{u=0}=0$ and $dR/du|_{u=0}=X_0$.

\begin{figure}[b!]
        \begin{center}
                \setlength{\tabcolsep}{ 20 pt }
                \begin{tabular}{cc}
                        \includegraphics[width=5cm]{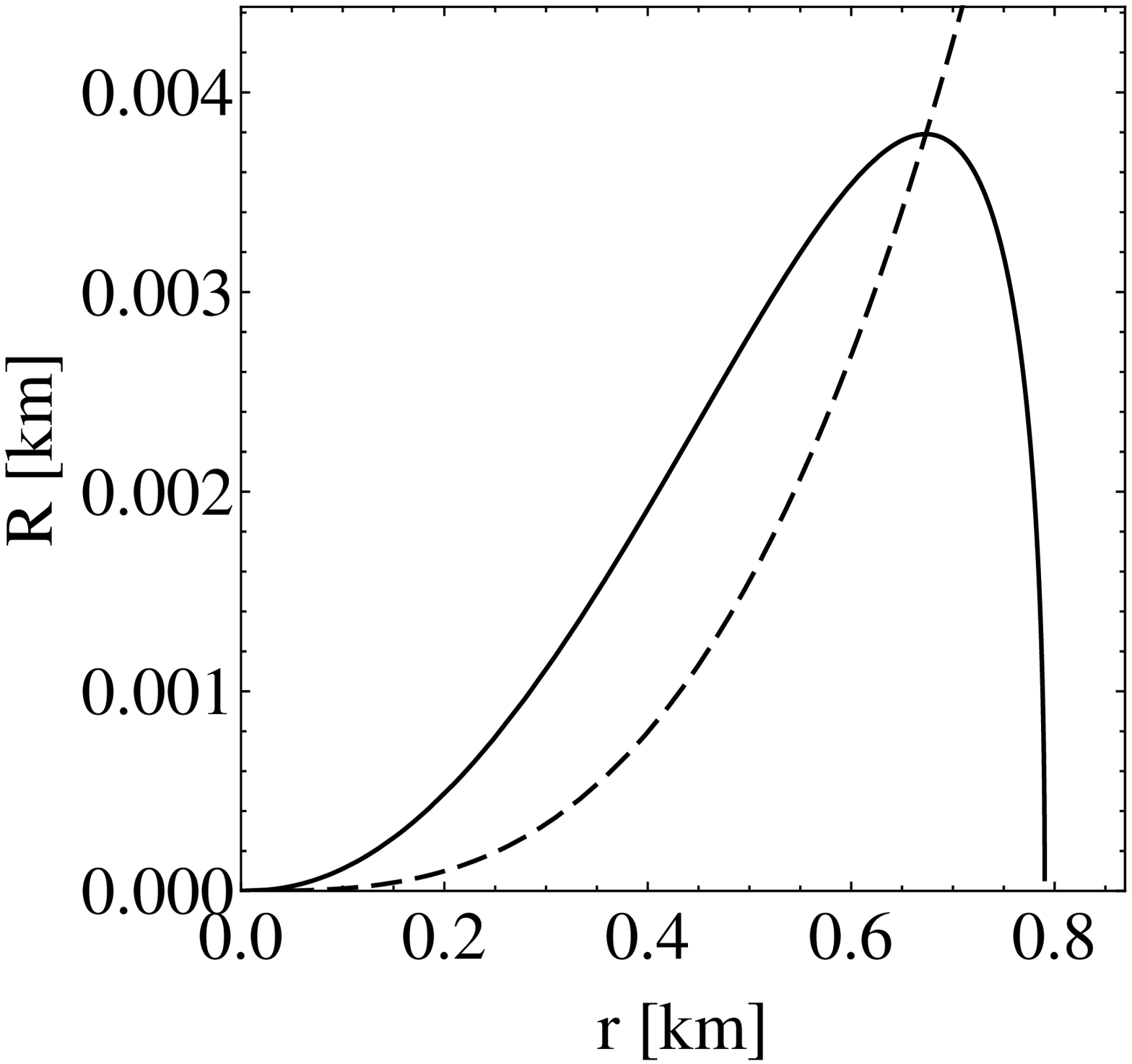} &
                        \includegraphics[width=5cm]{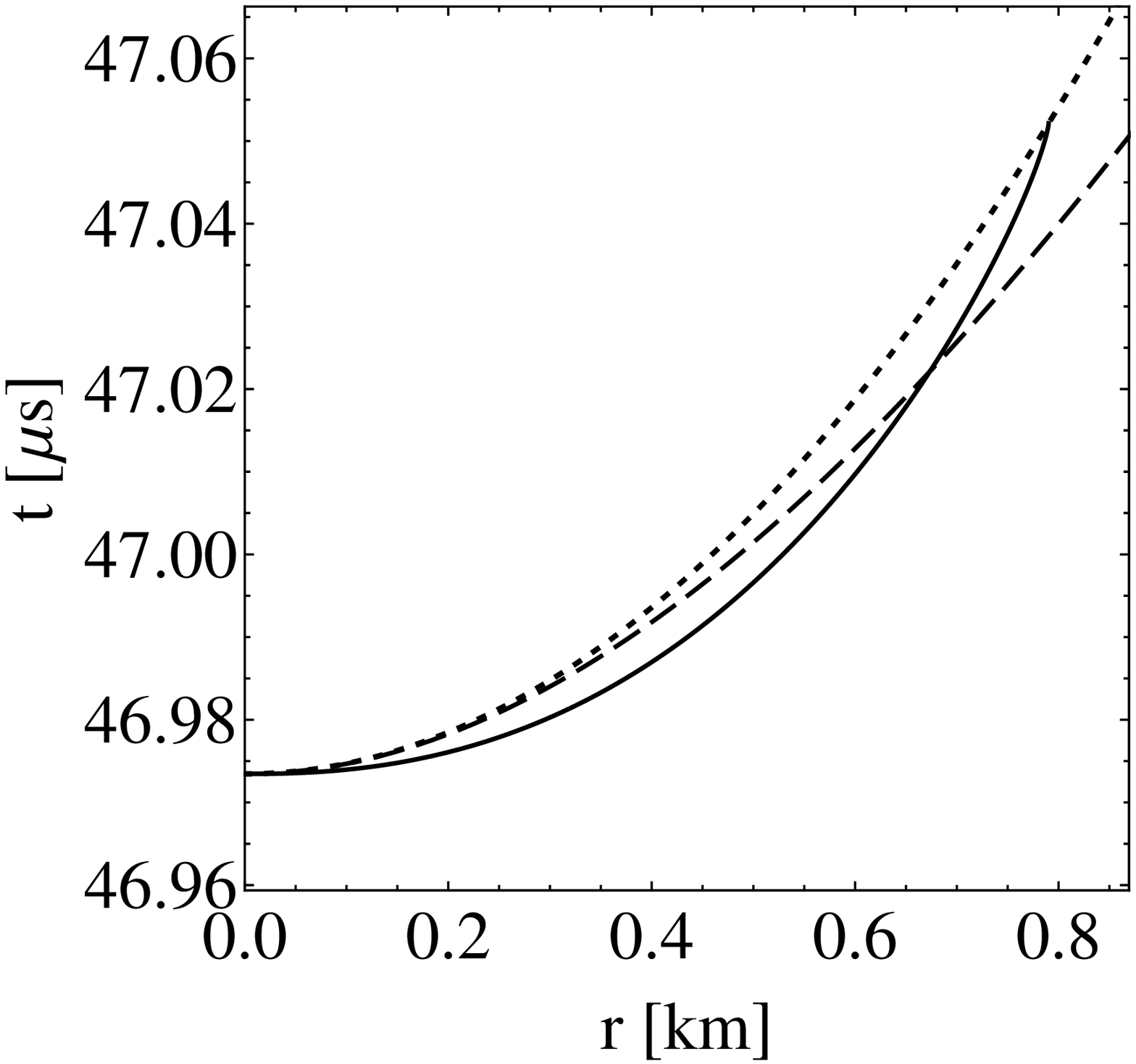} \\
                        (a) & (b) \\
                \end{tabular}
        \caption{(a) The trapping of the Cauchy horizon (solid) by the
	 apparent horizon (dashed) in the $r$-$R$ plane for the model B
	 with $n=2$. The Cauchy horizon emanating from the central
	 singularity $R=r=0$ is trapped by the apparent horizon at
	 $(r_{\rm trap},R_{\rm trap}) = (0.673 \; {\rm km}, 0.00379 \;{\rm
	 km})$, which is deep 
	 inside the stellar surface located at
	 $r=r_b = 10.5 \;{\rm km}$, and then plunges into the non-central
	 singularity lying on $R=0$ at $r=0.790 \;{\rm km}$. (b) The
	 trapping of the Cauchy horizon (solid) by the apparent horizon
	 (dashed) in the $r$-$t$ plane again for the model B with
	 $n=2$. The shell-focusing singularity is drawn by a dotted
	 curve, into which the Cauchy horizon plunges at $(r,t)=(0.790
	 \;{\rm km},47.052 \; \mu {\rm s})$.}
        \label{fg:B2}
        \end{center}
\end{figure}

\begin{figure}[tb]
        \begin{center}
        \includegraphics[width=5cm]{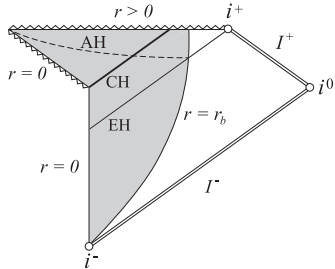}
        \caption{The conformal diagram of the LTB solution described in Sec.~\ref{sec:weak}. The central ($r=0$) singularity is locally naked but not globally. Namely, the Cauchy horizon (CH, thick solid) is trapped by the apparent horizon (AH, thin dashed) and plunges into the non-central ($r>0$) singularity without reaching the surface of collapsing star ($r=r_b$). The line EH (thin solid) represents the event horizon.}
\label{fg:penrose}
        \end{center}
\end{figure}

Let us focus on model B with $n=2$. The qualitative behavior of the
other models is the same as in this case. The Cauchy horizon numerically
obtained is shown in figure~\ref{fg:B2}. Figures~\ref{fg:B2}(a) and
\ref{fg:B2}(b) show the Cauchy horizon in the ($r,R$) and ($r,t$)
planes, respectively, with the apparent horizon \eqref{AH}. One can see
that $R$ increases with $r$ at first along the Cauchy horizon, but then
encounters the apparent horizon at $R=R_{\rm trap} = 3.79\;{\rm m}$ and
eventually plunges into the non-central singularity $R=0$ at $r >0 $. In
order to see that this trapping point is buried deep enough inside the
collapsing star, we define the areal radius of the collapsing star at the time of trapping by
\be
        R_b := R(t_{\rm trap},r_b),
\ee
where $t_{\rm trap}$ is the time of trapping measured in the
$t$-coordinate.\footnote{Although this areal radius has no
coordinate-independent meaning of course, it will do for our present
purpose.} In the present case, $R_b =7.56 \;{\rm km} $. Thus, $R_{\rm
trap}/R_b \simeq 0.05 \% $ and we can say that the trapping happens deep inside the star. In other words, the naked singularity cannot be globally naked, preserving the weak cosmic censorship hypothesis. A conformal diagram is depicted in figure~\ref{fg:penrose}. See table~\ref{tbl:para} for the values of parameters for the other models.

\subsection{How small mass is required to violate the weak CCH?}
\label{sec:violation}

We saw above that the naked singularities appearing in the collapse of
stellar-size objects cannot be globally naked. Here, we illustrate that
globally naked singularities can happen {\it but} only for parameters
that are unrealistic. Let $r_b$ be given by the right-hand side of
inequality~\eqref{constraint}, and we take $n=2$ and the value of
$\rho_c$ of model B. Then, only parameter left to be fixed is the mass
$M$. Integrating equation~\eqref{dRdu} numerically up to the surface for
various values of $M$, one can easily find that for $M 
\gtrsim 0.2466 M_\odot $ the Cauchy horizon is trapped by the apparent
horizon before reaching the surface of star as in the cases in
section~\ref{sec:weak}, whereas for $M 
\lesssim 0.2466 M_\odot$ the Cauchy horizon is not trapped within the stellar interior. Thus, the globally naked singularities can appear only for collapse of the extremely small-mass regime, that is far below the masses of the observed neutron stars ($\sim 1.5 M_{\odot}$) and the maximum mass $(\sim 1.5 M_{\odot})$ of the neutron star for the extremely soft equation of state.

\section{Conclusion}
\label{sec:conc}

In this paper, we have investigated the collapse of massive objects described by the Lema\^{i}tre-Tolman-Bondi solution~\eqref{metric}, which raised some of the most serious counterexamples to violate the strong cosmic censorship hypothesis. We impose the initial conditions characterized by the initial density profile \eqref{rho} and velocity distribution \eqref{f}, setting the initial values for the central density, mass, and radius to those of marginally stable neutron stars (see table \ref{tbl:para}). For $n=1$ and $n=2$, the singularity in all models is necessarily locally naked, but we have shown that it cannot be globally naked. Namely, the first null ray emanating from the singularity cannot reach the future null infinity $I^+$ (see figure \ref{fg:penrose}), implying that the singularity cannot be seen by any observers except those who inevitably plunge into the future spacelike singularity. Thus, it can be said that the singularity is censored by the astrophysical censor in the weak sense.

We would like to stress again that we are not in the position to claim that there appear locally naked singularities in the astrophysically realistic collapse. The appearance of naked singularities in our analysis is entirely because we choose the dust as a matter model and the situation may be very different for other choices. Instead, our conclusion is that {\it even if} a naked singularity could appear at the centre of spherical collapse in astrophysical situations, it will be already hidden behind an event horizon and cannot be observed by a distant observer, \ie, the singularity can be locally naked but not globally. In other words, the violation of the weak cosmic censorship hypothesis, which can provide us with the chance to observe trans-Planckian physics, seems unlikely at least in the spherically symmetric gravitational collapse of stars made of dust like matter.

\section*{Acknowledgments}

UM and TH are supported by Research Center for Measurement in Advanced Science in Rikkyo University. TH is supported by the Grant-in-Aid for Scientific Research Fund of the Ministry of Education, Culture, Sports, Science and Technology, Japan [Young Scientists (B) 21740190]. SJ acknowledges the support under the UGC minor research project on Black holes and Visible singularities.

\appendix

\section{The case of $n=3$}
\label{sec:strong}

In contrast to the cases of $n=1$ and $n=2$, where the central singularity is necessarily naked, the nakedness of the singularity for $n=3$ depends on the other parameters ($M,\rho_c,r_b$) (it turns out that $\alpha=3$ in this case). That is, the algebraic equation~\eqref{rootEq} has no positive root for certain parameter regions of ($M,\rho_c,r_b$), implying that the collapse results in a black hole. As will be shown below, one can see that for a typical set of ($M,\rho_c,r_b$) motivated by stellar collapse, algebraic equation~\eqref{rootEq} has no positive root.

In order to reduce the number of parameters and simplify our arguments, let $r_b$ be given by the right-hand side of the inequality~\eqref{constraint}, which results in $L=r_b$. Furthermore, we assume that $\rho_c$ takes the value of model B. Then, the only parameter left to be fixed is the mass $M$.

One can check numerically that the algebraic equation~\eqref{rootEq} has a positive root for $M \leq 0.0688 M_\odot$. Integrating equation~\eqref{dRdu} for a mass below this threshold, one can see also that the Cauchy horizon is never trapped inside the star ($M = 0.0688 M_\odot$ is a marginal case where the Cauchy horizon is trapped just at the surface), implying that the singularity is globally naked. On the other hand, for $M > 0.0688 M_\odot$ equation~\eqref{dRdu} has no positive root, implying the formation of a black hole. Thus, for the collapse of massive objects whose mass is a few solar mass the singularity cannot even be locally naked.



\end{document}